\documentclass[journal]{IEEEtran}
\usepackage{amsmath,amssymb,amsfonts}
\usepackage{booktabs}
\usepackage{algorithm}
\usepackage{algpseudocode}
\usepackage{array}
\usepackage[caption=false,font=footnotesize]{subfig}
\usepackage{graphicx}
\usepackage{multirow}
\usepackage{siunitx}
\usepackage{url}
\usepackage[hidelinks]{hyperref}
\usepackage{ifthen}
\usepackage{nomencl}
\makenomenclature
\begin{document}

\title{Hosting Capacity Assessment of Data Centers with Voltage Ride-Through Capability in Power Systems}

\author{Pengyu~Ren,~\IEEEmembership{Student Member,~IEEE,}
        , Wei~Sun\textsuperscript{*},~\IEEEmembership{Member,~IEEE}
        , Fei~Teng,~\IEEEmembership{Senior Member,~IEEE}

\thanks{Pengyu and Wei are from the School of Engineering, University of Edinburgh. Fei is from the Department of Electrical and Electronic Engineering at Imperial College London.}
%

\thanks{\textsuperscript{*}Corresponding author: Dr Wei Sun
(email: W.Sun@ed.ac.uk).}}

\maketitle

\begin{abstract}
Large data centers are emerging as concentrated, power-electronic grid loads whose abrupt disconnection or transfer to on-site backup supply during voltage disturbances can remove large demand from the power system, and may create a system-level stability problem. Their interconnection feasibility therefore depends not only on steady-state thermal and voltage limits, but also on whether internal power-conditioning systems can maintain IT service while limiting customer-initiated load reduction. This paper presents a voltage ride-through (VRT)-aware data center and grid co-planning framework that couples transmission-level fault simulation with an internal data center ride-through model. Python-based dynamic simulations generate point-of-interconnection (POI) voltage trajectories under selected network faults, and the resulting waveforms drive an internal model incorporating IT and cooling-load dynamics, DC-link, Uninterruptible Power Supply (UPS) response, and converter apparent power limits. The IEEE 118-bus case study shows that internal VRT capability can become a binding interconnection constraint: steady-state planning alone can overestimate feasible data center capacity, whereas increased UPS converter headroom progressively restores hosting capacity. Under the reduced-order response models studied, the grid-forming mode provides greater ride-through margin than the current-limited grid-following mode under the same network fault conditions. The results further show that VRT constraints can materially change both the total hosting capacity of data centers and its spatial allocation across candidate interconnection buses.
\end{abstract}


\IEEEpeerreviewmaketitle

\renewcommand{\nomgroup}[1]{%
  \item[\bfseries
  \ifthenelse{\equal{#1}{A}}{Abbreviations}{%
  \ifthenelse{\equal{#1}{S}}{Sets and Indices}{%
  \ifthenelse{\equal{#1}{V}}{Variables}{%
  \ifthenelse{\equal{#1}{P}}{Parameters}{}}}}%
]}

\nomenclature[A]{BESS}{Battery Energy Storage System}
\nomenclature[A]{DC}{Data Center}
\nomenclature[A]{GFL}{Grid-following response mode}
\nomenclature[A]{GFM}{Grid-forming response mode}
\nomenclature[A]{POI}{Point of Interconnection}
\nomenclature[A]{UPS}{Uninterruptible Power Supply}
\nomenclature[A]{VRT}{Voltage Ride-Through}

\nomenclature[S]{$\mathcal{N}$}{Set of buses}
\nomenclature[S]{$\mathcal{L}$}{Set of transmission branches}
\nomenclature[S]{$\mathcal{C}$}{Set of candidate data center buses}
\nomenclature[S]{$\mathcal{T}$}{Set of operating periods}
\nomenclature[S]{$\mathcal{T}^{\mathrm{VRT}}$}{Subset of operating periods used for internal VRT surrogate constraints}
\nomenclature[S]{$\mathcal{K}$}{Set of sampled VRT time points}
\nomenclature[S]{$\mathcal{F}$}{Set of tested network faults}
\nomenclature[S]{$i,j$}{Bus indices}
\nomenclature[S]{$t$}{Operating-period index}
\nomenclature[S]{$k$}{VRT sample-time index}
\nomenclature[S]{$f$}{Network fault index}
\nomenclature[S]{$\tau$}{Continuous post-disturbance time}

\nomenclature[V]{$H_i$}{Hosted data center capacity at bus \(i\) (MW)}
\nomenclature[V]{$P^{\mathrm{dc}}_{i,t},Q^{\mathrm{dc}}_{i,t}$}{Total data center active and reactive facility load at bus \(i\), period \(t\) (MW, MVAr)}
\nomenclature[V]{$P^{\mathrm{IT}}_{i,t},P^{\mathrm{cool}}_{i,t}$}{IT and cooling/auxiliary active load components (MW)}
\nomenclature[V]{$P^{\mathrm{g}}_{i,t}$}{Dispatchable generation or source injection at bus \(i\), period \(t\) (MW)}
\nomenclature[V]{$F_{ij,t}$}{Active power flow on branch \((i,j)\), period \(t\) (MW)}
\nomenclature[V]{$\theta_{i,t}$}{Voltage angle at bus \(i\), period \(t\) (rad)}
\nomenclature[V]{$V^{\mathrm{POI}}_{i,t,f}(\tau)$}{POI voltage trajectory at bus \(i\) under fault \(f\) (p.u.)}
\nomenclature[V]{$V^{\mathrm{IT}}_{i,t,f}(\tau)$}{Internal protected IT-bus voltage trajectory (p.u.)}
\nomenclature[V]{$\underline V^{\mathrm{IT},\Delta}_{i,t,f}$}{Worst \(\Delta\)-ms sustained internal IT voltage (p.u.)}
\nomenclature[V]{$L^{\mathrm{loss}}_{i,t,f}$}{Maximum POI active-load loss during fault \(f\) (MW)}
\nomenclature[V]{$L^{\mathrm{agg}}_{t,f}$}{Aggregate POI active-load loss across candidate data centers (MW)}
\nomenclature[V]{$\hat m^V_{i,t,f},\hat m^L_{i,t,f}$}{Surrogate margins for internal voltage and POI load loss}
\nomenclature[V]{$P^{\mathrm{grid}}_i(\tau),P^{\mathrm{BESS}}_i(\tau)$}{Protected-path active power supplied from the grid and BESS (MW)}
\nomenclature[V]{$e^{\mathrm{dc}}_i(\tau)$}{Normalized DC-link energy state}

\nomenclature[P]{$P^{\mathrm{base}}_{i,t}$}{Non-data-center active load at bus \(i\), period \(t\) (MW)}
\nomenclature[P]{$P^{\mathrm{w}}_{i,t}$}{Wind active power injection at bus \(i\), period \(t\) (MW)}
\nomenclature[P]{$V^{\mathrm{IT,req}}$}{Minimum acceptable internal IT voltage (p.u.)}
\nomenclature[P]{$d_t$}{Normalized data center load factor}
\nomenclature[P]{$\mathrm{pf}$}{Data center power factor}
\nomenclature[P]{$B_{ij}$}{Linearized branch susceptance coefficient}
\nomenclature[P]{$\bar S_{ij}$}{Branch active-flow limit (MW)}
\nomenclature[P]{$V^0_{\ell,t}$}{Base-case voltage magnitude at bus \(\ell\), period \(t\) (p.u.)}
\nomenclature[P]{$\Gamma_{\ell i,t}$}{AC voltage sensitivity of bus \(\ell\) to data center capacity at bus \(i\)}
\nomenclature[P]{$\underline V,\bar V$}{Normal-operation voltage magnitude bounds (p.u.)}
\nomenclature[P]{$s_{\max}$}{Branch-rating scale factor in the case study}
\nomenclature[P]{$\mathrm{PUE}$}{Data center power usage effectiveness}
\nomenclature[P]{$S^{\mathrm{rated}}_i$}{Data center UPS/BESS converter apparent-power rating (MVA)}
\nomenclature[P]{$\beta$}{UPS/BESS converter rating ratio \(S^{\mathrm{rated}}/H\)}
\nomenclature[P]{$E^{\mathrm{dc}}_i$}{Available DC-link hold-up energy (MWs)}
\nomenclature[P]{$\bar P^{\mathrm{BESS}}_i,\bar E^{\mathrm{BESS}}_i$}{BESS active-power and energy limits (MW, MWh)}
\nomenclature[P]{$I^{\max}_i$}{Fault-period POI current limit for the protected path}
\nomenclature[P]{$\lambda^{\mathrm{loss}}$}{Maximum allowed POI load-loss fraction}
\nomenclature[P]{$\lambda^{\mathrm{agg}}$}{Maximum allowed aggregate POI load-loss fraction}
\nomenclature[P]{$\tilde a^q_{i,f},b^q_{i,j,f},c^q_{i,f}$}{Tightened affine surrogate coefficients for margin \(q\)}

\printnomenclature[1.1in]

\section{Introduction}
Large data centers are becoming one of the most consequential sources of new electric demand. Their growth is driven by cloud services, high-density computing, and artificial-intelligence workloads. The International Energy Agency (IEA) projects that data center electricity consumption will roughly double from \SI{485}{TWh} in 2025 to \SI{950}{TWh} in 2030, accounting for around \SI{3}{\percent} of global electricity demand by that date \cite{iea2026keyquestionsai}. This growth is already visible in interconnection processes: an ERCOT interconnection update reported approximately \SI{410}{GW} of large loads seeking interconnection as of March 26, 2026, of which about \SI{87}{\percent} were data centers \cite{ercot2026interconnectionupdate}. The central planning question is therefore no longer only how much energy data centers will consume, but how large data centers can be connected without creating unacceptable dynamic reliability risk.

The difficulty is that modern data centers are not static loads. They are power-electronic facilities with rectifiers, UPS systems, battery energy storage, cooling systems, server power supplies, and increasingly sophisticated inverter controls. During a voltage disturbance, a data center may remain grid-connected, partially reduce power drawn from the grid, transfer part of its demand to internal storage, or disconnect from the system to protect IT service. This behavior matters at transmission scale. An NERC incident review reported that a 230-kV transmission-line fault caused approximately 1500 {MW} of customer-initiated voltage-sensitive load reduction, largely associated with data-center-type loads, even though the fault was cleared by utility protection \cite{nerc2025largeloadloss}. PJM Interconnection (PJM) subsequently documented a separate July 2026 event in which approximately \SI{3800}{MW} transferred to backup power following a 230-kV line fault \cite{pjm2026dominion}. Such events expose a tension at the data center--grid interface: the grid needs large loads to avoid abrupt simultaneous disconnection during normally cleared faults, whereas the facility must maintain high-quality power for IT and cooling service.

This concern has led system operators to pay closer attention to voltage ride-through (VRT) and dynamic model quality requirements for large loads. Voltage ride-through requirements are well established for inverter-based resources, including IEEE Std 1547-2018 \cite{ieee1547}, and system operators specify voltage-versus-time envelopes for disturbance performance \cite{ercotvrt}. For large electronic loads, however, the relevant question is not simply whether the point-of-interconnection (POI) voltage is restored above a prescribed envelope. ERCOT's PGRR144 material explicitly frames a large-load model-quality-test waveform for assessing how large-load dynamic models respond to voltage ramp, shallow-fault, deep-fault, and high-voltage events \cite{ercot2026pgrr144}. Recent work has also examined data center low-voltage ride-through and internal voltage control \cite{xie2025datacenterlvrt}, grid-forming BESS as line-interactive UPS infrastructure \cite{azizi2026gfmups}, centralized UPS control modes for fault buffering and post-fault return \cite{shamseldein2025threemode}, active/reactive power limits for voltage ride-through in large loads \cite{norouzi2026dualpq}, and the role of current limiting and transient synchronization in grid-forming inverter ride-through \cite{ordono2024currentlimiting}. Together, these developments indicate that data center interconnection planning must consider the facility-side voltage ride-through response.

\begin{figure*}[t]
\centering
\includegraphics[width=1.0\textwidth]{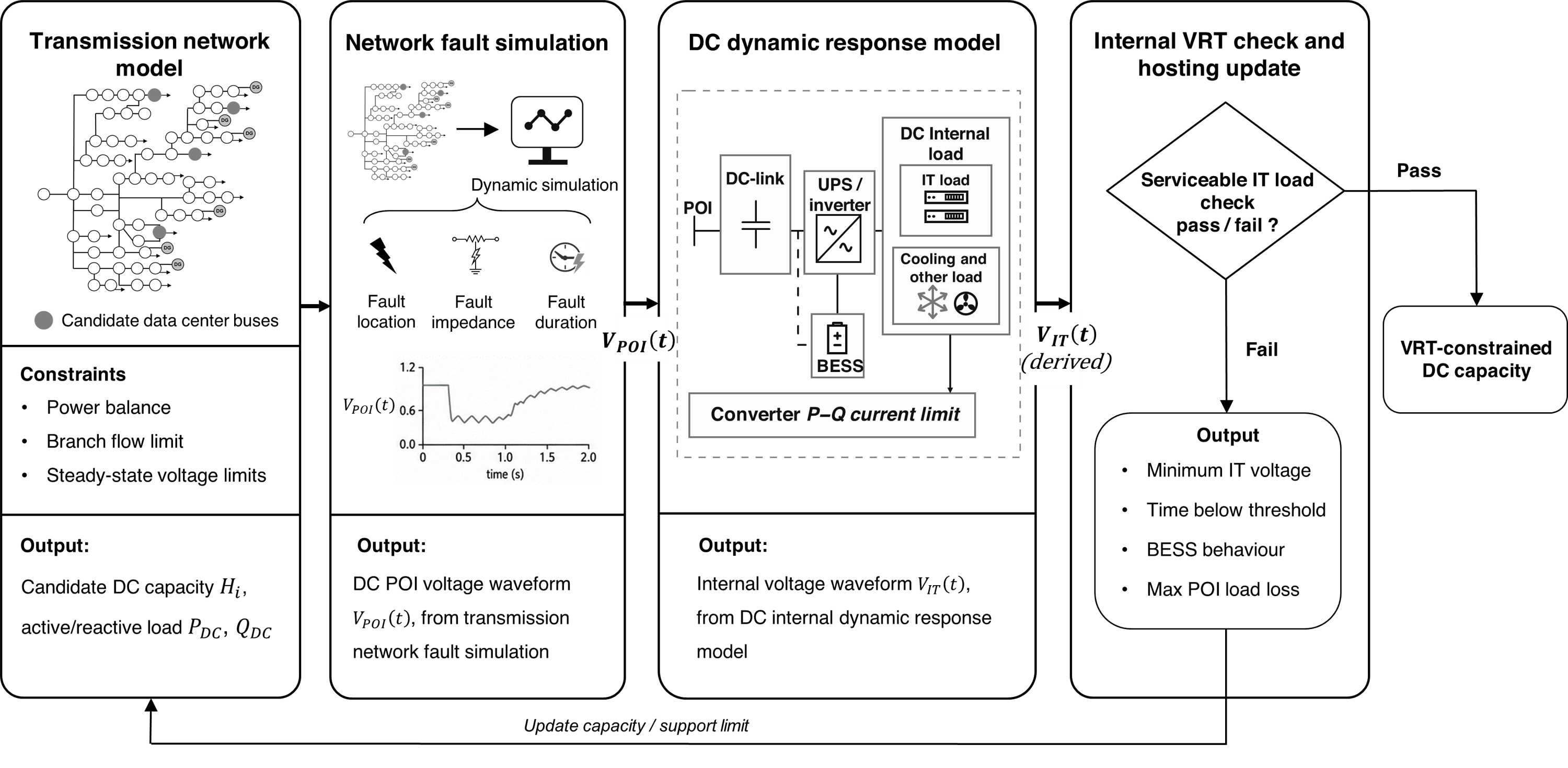}
\caption{Overall framework for VRT-aware data center--grid co-planning. A hosting proposal is converted into network fault simulations, POI voltage waveforms are passed into the internal data center dynamic model, and the resulting IT-voltage response determines whether the candidate capacity is accepted or updated.}
\label{fig:framework}
\end{figure*}

At the same time, most power-system planning models for data centers remain dominated by steady-state or quasi-static representations. Existing studies have considered expansion planning across electricity and data networks \cite{vafamehr2019}, spatial load migration and market participation by geographically distributed data centers \cite{zhang2020virtual,liu2021adn}, large-load integration and flexible-load potential in power systems \cite{duke2025largeflex}, and demand response or multi-energy operation \cite{caprino2019,ghadi2020}. Classical hosting-capacity and network-screening methods likewise focus on power balance, voltage magnitude, branch loading, and source limits \cite{epri_hosting,farivar2013,bolognani2016}. These models are essential for identifying where large loads may connect under normal operation, but they generally do not ask whether the accepted data center capacity can ride through a transmission fault without unacceptable internal voltage degradation or abrupt load loss.

Despite progress in both areas, a gap remains between facility-level data center ride-through studies and transmission-level interconnection planning. Data center VRT studies typically use prescribed voltage sags or detailed facility test systems, while planning studies generally do not feed network-dependent fault voltage waveforms into an internal data center response model. Conversely, a steady-state interconnection solution can identify a feasible MW allocation, but it cannot determine whether the resulting POI voltage trajectory during a selected fault is compatible with UPS dynamics, DC-link hold-up, BESS support, converter apparent-power limits, IT load dynamics, and cooling-load response. The missing link is a planning framework in which data center VRT capability changes the acceptable amount and location of data center interconnection through a network-coupled dynamic check.

To the best of our knowledge, our work is the first to couple transmission-level data center capacity planning with an internal data center voltage ride-through response model driven by fault-dependent POI voltage waveforms. The main contributions are:

\begin{itemize}
    \item We formulate a VRT-aware data center planning problem that treats large data center interconnection as a joint question of network hosting feasibility and facility-level ride-through capability, rather than as a steady-state load-connection problem alone.
    \item Instead of representing data centers as static aggregate loads or imposing ride-through feasibility directly at the POI, we developed a dynamic response model for VRT-capable data centers. The model captures UPS dynamics, DC-link hold-up, BESS support, IT and cooling load behavior, converter limits, and GFL/GFM response modes, and is coupled with fault simulations to embed facility-level ride-through behavior into power-system planning.
    \item We validate the framework on an IEEE 118-bus case study and show that facility-level VRT capability can materially change the feasible interconnection capacity and its spatial allocation. Under the reduced-order response models used here, the results reveal a monotonic recovery of hosting capacity as current-limited GFL support is strengthened, with the GFM proxy providing a higher ride-through margin under the same network fault conditions.
\end{itemize}

\section{Methodology}
Fig.~\ref{fig:framework} summarizes the proposed framework. The method is organized as a sequential data center--grid co-planning loop. First, a steady-state hosting problem proposes candidate data center capacities and hourly operating points. Second, for each proposed hosting point, transmission-network fault simulations generate POI voltage waveforms at the candidate data center buses. Third, those POI waveforms drive a reduced-order data center dynamic response model with DC-link, UPS output-stage, BESS, IT-load, cooling-load, and converter-limit components. Fourth, the internal VRT checker evaluates whether the protected IT load remains serviceable and returns pass/fail labels and diagnostics. Failed hosting points are reduced or re-tested with different ride-through capability assumptions until a VRT-constrained capacity is obtained.

\subsection{Problem Formulation}
Let $\mathcal{N}$ be the bus set, $\mathcal{L}$ the directed branch set, $\mathcal{T}$ the operating-period set, $\mathcal{F}$ the set of selected network faults, and $\mathcal{C}\subseteq\mathcal{N}$ the candidate data center bus set. The planning variable $H_i$ denotes the total facility-load capacity accepted at candidate bus $i$, not only the IT load. The steady-state hosting proposal maximizes accepted data center capacity,
\begin{equation}
\max \sum_{i\in\mathcal{C}} H_i ,
\label{eq:objective}
\end{equation}
subject to normal-operation network constraints. The VRT-aware framework then screens the proposed capacity with dynamic fault simulations and the internal data center response model.

The active data center demand associated with an installed capacity is
\begin{equation}
P^{\mathrm{dc}}_{i,t}=d_t H_i,\quad i\in\mathcal{C},\;t\in\mathcal{T},
\label{eq:dc_profile}
\end{equation}
where $d_t$ is the normalized operating profile. The corresponding reactive demand is represented through an assumed power factor,
\begin{equation}
Q^{\mathrm{dc}}_{i,t}
=P^{\mathrm{dc}}_{i,t}\tan(\arccos(\mathrm{pf})).
\label{eq:dc_q}
\end{equation}

\subsection{Network Constraints}
The hosting model uses a linearized transmission network model. For each directed branch $(i,j)\in\mathcal{L}$ and operating point $t$, active power flow is represented by
\begin{equation}
F_{ij,t}=B_{ij}(\theta_{i,t}-\theta_{j,t}),
\label{eq:dcflow}
\end{equation}
where $B_{ij}$ is the branch susceptance coefficient and $\theta_{i,t}$ is the bus voltage angle. Branch thermal limits are enforced as
\begin{equation}
-\bar S_{ij}\leq F_{ij,t}\leq \bar S_{ij}.
\label{eq:linelimit}
\end{equation}
Power balance is enforced at every bus and operating point:
\begin{equation}
\sum_{j:(i,j)\in\mathcal{L}}F_{ij,t}-\sum_{j:(j,i)\in\mathcal{L}}F_{ji,t}
+P^{\mathrm{base}}_{i,t}+P^{\mathrm{dc}}_{i,t}-P^{\mathrm{w}}_{i,t}-P^{\mathrm{g}}_{i,t}=0 .
\label{eq:balance}
\end{equation}
Dispatchable source output, candidate capacity, and voltage-angle variables satisfy
\begin{align}
\underline P^{\mathrm{g}}_{i,t}\leq P^{\mathrm{g}}_{i,t}\leq \bar P^{\mathrm{g}}_{i,t},
\quad i\in\mathcal{N},\;t\in\mathcal{T},
\label{eq:gen_bounds}\\
0\leq H_i\leq \bar H_i,
\quad i\in\mathcal{C},
\label{eq:h_bounds}\\
\theta_{r,t}=0,
\quad t\in\mathcal{T},
\label{eq:slack_angle}
\end{align}
where $r$ is the reference bus. To avoid accepting a hosting point that is feasible only under a linearized active-power approximation, the planning model includes a linearized AC-voltage screening constraint. Around a base AC operating point,
\begin{equation}
V^{\mathrm{op}}_{\ell,t}
=
V^0_{\ell,t}
+
\sum_{i\in\mathcal{C}}\Gamma_{\ell i,t}H_i ,
\quad \ell\in\mathcal{N},\;t\in\mathcal{T},
\label{eq:voltage_proxy}
\end{equation}
where $\Gamma_{\ell i,t}$ is the voltage sensitivity to data center capacity at candidate bus $i$. The normal-operation voltage screen is
\begin{equation}
\underline V
\leq
V^{\mathrm{op}}_{\ell,t}
\leq
\bar V,
\quad \ell\in\mathcal{N},\;t\in\mathcal{T}.
\label{eq:voltage_proxy_limit}
\end{equation}
The final optimized points are also checked by a full AC power flow in the validation stage; the linear proxy is used only to keep the hosting search in a normal-voltage region before dynamic fault checking.

\subsection{Data Center Modeling}
The data center is represented at two levels. In the hosting proposal it is a controllable siting-and-sizing decision through $H_i$ and $P^{\mathrm{dc}}_{i,t}$. In the dynamic checker it is a facility-level response model driven by the POI voltage waveform. The dynamic model separates protected IT demand, cooling demand, auxiliary demand, DC-link energy, UPS output-stage voltage control, and BESS support. The BESS is modeled as a DC-side energy source connected to the UPS DC link; it supplies active-power deficit during ride-through, while the UPS output-stage converter regulates the protected internal AC bus subject to the same apparent-power headroom.

\begin{figure}[t]
\centering
\includegraphics[width=0.9\linewidth]{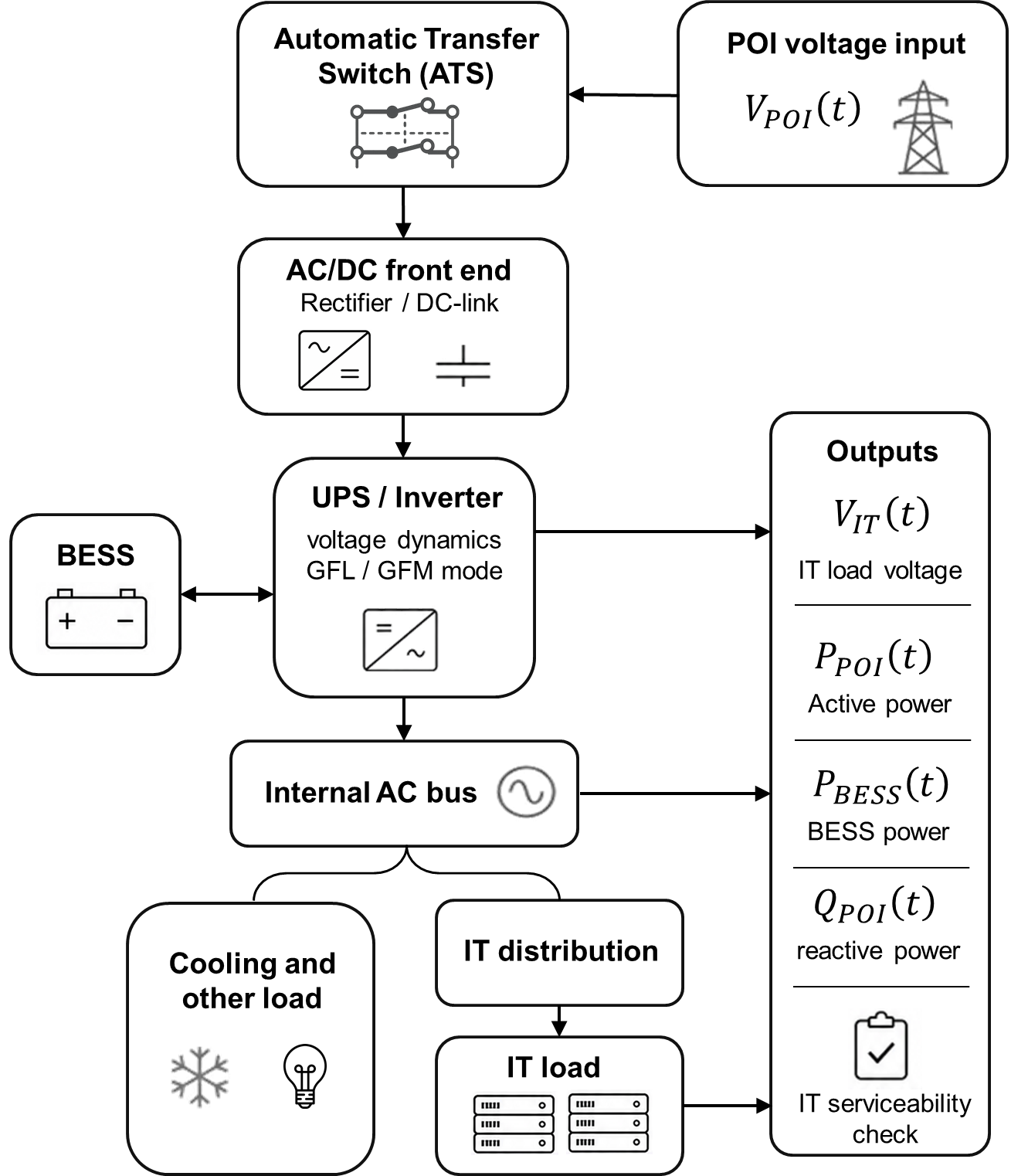}
\caption{Internal data center model for VRT assessment. The POI voltage drives a rectifier/DC-link stage; the DC-side BESS supplies ride-through energy; the UPS output-stage converter regulates the protected IT bus while serving IT, cooling, and auxiliary loads under a shared apparent-power limit.}
\label{fig_internal_DC}
\end{figure}

\subsubsection{IT and Cooling Load Modeling}

The protected IT component can be represented as the superposition of workload components,
\begin{equation}
P^{\mathrm{IT}}_i(\tau)
=
P^{\mathrm{tr},0}_i(\tau)
+\sum_{j\in\mathcal{J}^{\mathrm{tr}}}P^{\mathrm{tr},j}_i(\tau)
+\sum_{j\in\mathcal{J}^{\mathrm{ft}}}P^{\mathrm{ft},j}_i(\tau).
\label{eq:ai_workload}
\end{equation}
Each component alternates between high-compute and lower communication or synchronization phases, with stochastic cycle durations, phase lengths, and short-term perturbations. A gain parameter scales the fluctuation amplitude around the mean load, so the same structure can represent either highly synchronized AI computation or partially smoothed operation.

Cooling and auxiliary loads affect converter headroom and POI active power. Let $g^{\mathrm{IT}}_i(\tau)$ be the normalized IT load multiplier. The cooling multiplier $g^{\mathrm{cool}}_i(\tau)$ is modeled as a first-order response to IT-power variation,
\begin{equation}
T^{\mathrm{cool}}\dot g^{\mathrm{cool}}_i
=
1+\rho^{\mathrm{cool}}\left(g^{\mathrm{IT}}_i-1\right)
-g^{\mathrm{cool}}_i ,
\label{eq:cooling_response}
\end{equation}
where $T^{\mathrm{cool}}$ is the cooling thermal time constant and $\rho^{\mathrm{cool}}$ determines how strongly cooling power follows IT-power variation. Setting $\rho^{\mathrm{cool}}=0$ recovers a quasi-constant non-server-load assumption. The instantaneous facility load passed to the internal VRT model is
\begin{equation}
P^{\mathrm{dc}}_i(\tau)=P^{\mathrm{IT}}_i(\tau)+P^{\mathrm{cool}}_i(\tau)+P^{\mathrm{aux}}_i .
\label{eq:dc_dynamic_load}
\end{equation}
This distinction matters because the UPS/BESS primarily protects IT service, while cooling and auxiliary load still affect POI power and converter headroom.

\subsubsection{UPS Modelling}

\begin{figure*}[t]
\centering
\includegraphics[width=0.9\textwidth]
{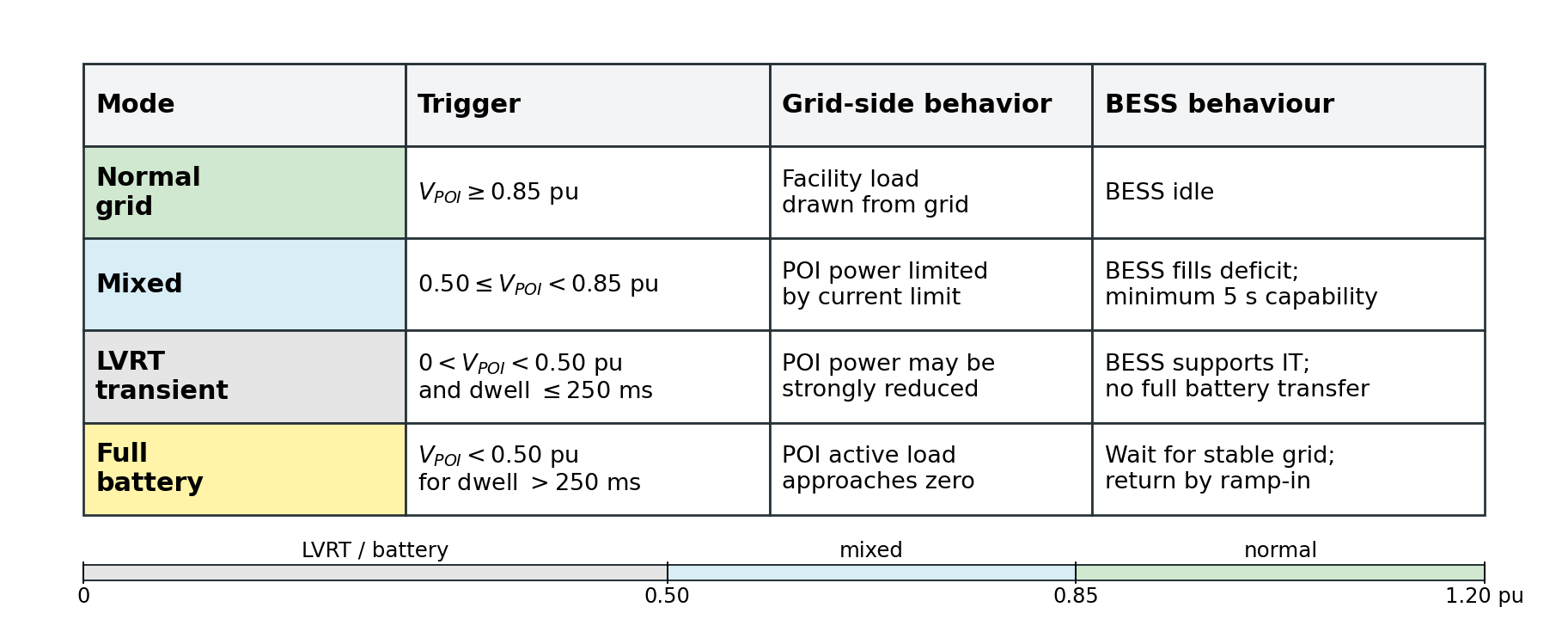}
\caption{Four-region UPS supervisory logic and staged reconnection in this model.}
\label{fig:ups_modes}
\end{figure*}

The voltage thresholds and dwell-time logic in
Fig.~\ref{fig:ups_modes} define the supervisory operating state of the
UPS \cite{entsoe2026datacentres}. During normal and mixed operation,
the facility remains electrically coupled to the grid, with the BESS
supporting the protected DC-side demand when the grid-side power is
insufficient. A persistent or deeper voltage depression initiates
battery operation, for which the POI active-power import approaches
zero, followed by controlled reconnection after the grid has remained
stable.

The grid-connected UPS front end is represented using either a
grid-following (GFL) or grid-forming (GFM) controller. Let
$\underline V_i$ denote the positive-sequence POI voltage and
$\underline I_i$ the current flowing from the grid into the data
centre. The import-positive power convention is
\begin{equation}
P_i^{\mathrm g}+jQ_i^{\mathrm g}
=
S_{\mathrm B}\underline V_i\underline I_i^{*},
\label{eq:ups_complex_power}
\end{equation}
where positive $P_i^{\mathrm g}$ and $Q_i^{\mathrm g}$ denote active-
and reactive-power import, respectively.

Both controller modes use a common DC-energy outer loop. Defining
\begin{equation}
e_i^{\mathrm{dc}}
=
\frac{E_i^{\mathrm{dc},\star}-E_i^{\mathrm{dc}}}
     {E_i^{\mathrm{dc},\star}},
\qquad
\dot{\xi}_i^{\mathrm{dc}}=e_i^{\mathrm{dc}},
\label{eq:ups_dc_error}
\end{equation}
the active- and reactive-power commands are
\begin{align}
P_i^{\mathrm c}
&=
\operatorname{sat}_{[P_i^{\min},P_i^{\max}]}
\left(
P_i^{\mathrm{ref}}
+
K_{p,\mathrm{dc}}e_i^{\mathrm{dc}}
+
K_{i,\mathrm{dc}}\xi_i^{\mathrm{dc}}
\right),
\label{eq:ups_p_command}\\
Q_i^{\mathrm c}
&=
\operatorname{sat}_{[Q_i^{\min},Q_i^{\max}]}
\left(Q_i^{\mathrm{ref}}\right).
\label{eq:ups_q_command}
\end{align}
The BESS supervisor adjusts the DC-side support according to the
DC-voltage state and the grid--load power mismatch, subject to energy
and state-of-charge limits. Conditional anti-windup prevents the
DC-energy controller from integrating an unavailable active-power
request.

For the GFL mode, the POI voltage is expressed in the PLL reference
frame as
\begin{equation}
\underline V_{i,dq}
=
\underline V_i e^{-j\delta_{i,\mathrm{pll}}},
\qquad
e_{i,\mathrm{pll}}
=
\operatorname{Im}\{\underline V_{i,dq}\}.
\label{eq:gfl_pll_error}
\end{equation}
The synchronous-reference-frame PLL is represented by
\begin{align}
\dot{\xi}_{i,\mathrm{pll}}
&=e_{i,\mathrm{pll}},\\
\omega_{i,\mathrm{pll}}
&=
\omega_0
+
K_{p,\mathrm{pll}}e_{i,\mathrm{pll}}
+
K_{i,\mathrm{pll}}\xi_{i,\mathrm{pll}},\\
\dot{\delta}_{i,\mathrm{pll}}
&=
\omega_{i,\mathrm{pll}}-\omega_0 .
\label{eq:gfl_pll}
\end{align}
The commanded complex current is
\begin{equation}
\underline I_{i,dq}^{\star}
=
\mathcal L^{P}_{I_i^{\max}}
\left[
\left(
\frac{P_i^{\mathrm c}+jQ_i^{\mathrm c}}
     {S_{\mathrm B}\underline V_{i,dq,\epsilon}}
\right)^{*}
\right],
\label{eq:gfl_current_reference}
\end{equation}
where $\underline V_{i,dq,\epsilon}$ applies a lower voltage bound to
avoid an unbounded current reference during deep voltage depressions.
The current controller is approximated by
\begin{equation}
\tau_i^{I}
\dot{\underline I}_{i,dq}
=
\underline I_{i,dq}^{\star}
-
\underline I_{i,dq},
\qquad
\underline I_i
=
\underline I_{i,dq}e^{j\delta_{i,\mathrm{pll}}}.
\label{eq:gfl_current_loop}
\end{equation}

For the GFM mode, the UPS is represented as an internal voltage source
behind a virtual impedance:
\begin{equation}
\underline E_i
=
E_i e^{j\delta_i},
\qquad
Z_i^{\mathrm v}
=
R_i^{\mathrm v}+jX_i^{\mathrm v}.
\label{eq:gfm_internal_voltage}
\end{equation}
Under the import-positive convention, the unconstrained and achieved
POI currents are
\begin{equation}
\underline I_i^{\mathrm u}
=
\frac{\underline V_i-\underline E_i}
     {Z_i^{\mathrm v}},
\qquad
\underline I_i
=
\mathcal L^{P}_{I_i^{\max}}
\left(\underline I_i^{\mathrm u}\right).
\label{eq:gfm_virtual_impedance}
\end{equation}
The internal frequency and angle follow virtual synchronous machine (VSM)
dynamics,
\begin{align}
2H_i^{\mathrm v}\dot{\omega}_i
&=
\frac{P_i^{\mathrm g}-P_i^{\mathrm{eff},\star}}
     {S_{\mathrm B}}
-
D_i^{\omega}(\omega_i-1),
\label{eq:gfm_vsm_frequency}\\
\dot{\delta}_i
&=
\omega_{\mathrm B}(\omega_i-1)
+
u_i^{\mathrm{sync}},
\label{eq:gfm_vsm_angle}
\end{align}
where $u_i^{\mathrm{sync}}$ is activated during controlled
resynchronization. The sign in
\eqref{eq:gfm_vsm_frequency} follows from defining positive power as
facility import rather than generator injection.

The internal-voltage magnitude is controlled through reactive-power
droop,
\begin{align}
E_i^{\star}
&=
\operatorname{sat}_{[E_i^{\min},E_i^{\max}]}
\left[
E_i^0
+
K_i^Q
\frac{Q_i^{\mathrm g}-Q_i^{\mathrm{ref}}}
     {S_{\mathrm B}}
\right],
\label{eq:gfm_voltage_command}\\
\tau_i^{E}\dot E_i
&=
E_i^{\star}-E_i .
\label{eq:gfm_voltage_dynamics}
\end{align}
When the AC-current limit, DC-power limit or resynchronisation logic is
active, power-matching anti-windup replaces the unavailable active-power
command by the achieved POI power:
\begin{equation}
P_i^{\mathrm{eff},\star}
=
\begin{cases}
P_i^{\mathrm c},
& \text{normal operation},\\[2mm]
P_i^{\mathrm g},
& \text{current/DC limitation or resynchronisation}.
\end{cases}
\label{eq:gfm_power_matching}
\end{equation}

The operator $\mathcal L^{P}_{I_i^{\max}}(\cdot)$ denotes an
active-power-priority current limiter. It enforces
\begin{equation}
|\underline I_i|
\leq I_i^{\max},
\qquad
\left(P_i^{\mathrm g}\right)^2
+
\left(Q_i^{\mathrm g}\right)^2
\leq
\left(
S_{\mathrm B}
|\underline V_i|
I_i^{\max}
\right)^2 ,
\label{eq:ups_current_limit}
\end{equation}
so reactive support is reduced when necessary to preserve active-power
service. The converter rating is parameterised as
\begin{equation}
S_i^{\mathrm{rated}}
=
\beta H_i,
\qquad
I_i^{\max}
=
\frac{S_i^{\mathrm{rated}}}{S_{\mathrm B}},
\label{eq:ups_rating_ratio}
\end{equation}
where $\beta$ represents the installed UPS/inverter headroom per unit
of hosted data-centre capacity.

The two modes therefore share the same DC-energy supervision, BESS
support and current-limiting layers, but differ in their grid-facing
control structure. The GFL mode follows the POI angle and regulates
current, whereas the GFM mode establishes its own voltage and frequency
through the virtual-impedance and VSM dynamics.

\subsection{Voltage Ride-Through Modeling}
The external grid-code VRT curve is not imposed as a required internal data center voltage trajectory. Instead, the transmission-network dynamic simulation provides the POI voltage waveform that the data center experiences during a fault. For a candidate hosting solution, fault $f$, and operating point $t$, the network simulation produces
\begin{equation}
V^{\mathrm{POI}}_{i,t,f}(\tau),\quad i\in\mathcal{C}.
\label{eq:poi_waveform}
\end{equation}
The dependence of this waveform on $H_i$ is the key coupling between hosting capacity and VRT. A larger hosted data center changes the pre-fault operating point through $P^{\mathrm{dc}}_{i,t}$ and $Q^{\mathrm{dc}}_{i,t}$, which can lower local voltage, consume reactive support headroom, and alter the post-fault voltage recovery under the same remote fault. Therefore the dynamic check is not performed with a fixed prescribed voltage sag for all capacities. For each tested hosting vector, the network case is rebuilt and the fault simulation is rerun so that
\begin{equation}
V^{\mathrm{POI}}_{i,t,f}(\tau)
=
\mathcal{G}_{\mathrm{grid}}
\left(
H,
d_t,
P^{\mathrm{w}}_t,
f
\right),
\label{eq:grid_fault_map}
\end{equation}
where $\mathcal{G}_{\mathrm{grid}}(\cdot)$ denotes the transmission-network power-flow and dynamic-fault simulation. This prevents the VRT check from becoming a stand-alone equipment test disconnected from the grid-planning decision.

The POI waveform is then used as the input to the internal data center model described above. The resulting protected IT-bus voltage is
\begin{equation}
V^{\mathrm{IT}}_{i,t,f}(\tau)
=
\mathcal{G}_{\mathrm{dc}}
\left(
V^{\mathrm{POI}}_{i,t,f}(\tau),
P^{\mathrm{IT}}_{i,t},
S^{\mathrm{rated}}_i,
E^{\mathrm{dc}}_i
\right),
\label{eq:internal_model_map}
\end{equation}
where $\mathcal{G}_{\mathrm{dc}}(\cdot)$ denotes the reduced-order DC-link, UPS, BESS, and converter-limit model.

The ride-through condition is based on sustained internal voltage. Let $\Delta$ be the voltage-duration window.
\begin{equation}
\underline V^{\mathrm{IT},\Delta}_{i,t,f}
=
\min_s
\left[
\min_{\tau\in[s,s+\Delta]}
V^{\mathrm{IT}}_{i,t,f}(\tau)
\right].
\label{eq:sustained_it_voltage}
\end{equation}

The hosting point is accepted only if
\begin{equation}
\underline V^{\mathrm{IT},\Delta}_{i,t,f}
\geq
V^{\mathrm{IT,req}},
\quad
i\in\mathcal{C},\;t\in\mathcal{T},\;f\in\mathcal{F}.
\label{eq:internal_vrt_constraint}
\end{equation}
The abrupt reduction in active power observed at the POI must also remain below a specified fraction of the data center operating load,
\begin{equation}
L^{\mathrm{loss}}_{i,t,f}
\leq
\lambda^{\mathrm{loss}} P^{\mathrm{dc}}_{i,t},
\quad
i\in\mathcal{C},\;t\in\mathcal{T},\;f\in\mathcal{F}.
\label{eq:load_loss_constraint}
\end{equation}

The first condition reflects whether the protected IT load remains serviceable, rather than whether the internal voltage is forced to stay above the external POI VRT envelope. The second condition limits the customer-initiated load loss at an individual site. Because simultaneous large-load loss is a system-level reliability concern, the checker also records the aggregate loss under each fault,
\begin{equation}
L^{\mathrm{agg}}_{t,f}
=
\sum_{i\in\mathcal{C}}L^{\mathrm{loss}}_{i,t,f}.
\label{eq:aggregate_loss}
\end{equation}
When a system-level load-loss requirement is specified, it can be imposed as
\begin{equation}
L^{\mathrm{agg}}_{t,f}
\leq
\lambda^{\mathrm{agg}}
\sum_{i\in\mathcal{C}}P^{\mathrm{dc}}_{i,t},
\quad t\in\mathcal{T},\;f\in\mathcal{F}.
\label{eq:aggregate_loss_constraint}
\end{equation}
In the case study, aggregate loss is reported as a diagnostic and used in sensitivity checks because a universal planning threshold depends on the balancing area and operating reserve context.

\subsection{Incorporating VRT constraints into Hosting Assessment}
Directly embedding the network fault simulation and internal data center dynamics inside the hosting optimization is not computationally practical. The implementation therefore samples the coupled network--data-center response offline and fits an optimization-ready linear surrogate. For each sampled hosting vector, the network case is rebuilt, a selected fault is simulated, and the resulting POI waveform is passed into the internal data center model. The checker returns two margins:
\begin{equation}
m^V_{i,t,f}
=
\underline V^{\mathrm{IT},\Delta}_{i,t,f}
-V^{\mathrm{IT,req}},
\label{eq:voltage_margin}
\end{equation}
and
\begin{equation}
m^L_{i,t,f}
=
\lambda^{\mathrm{loss}}P^{\mathrm{dc}}_{i,t}
-L^{\mathrm{loss}}_{i,t,f}.
\label{eq:loss_margin}
\end{equation}
The first margin is positive when the internal sustained-voltage criterion is satisfied; the second is positive when the POI load-loss limit is satisfied. The checker also records diagnostic quantities including instantaneous minimum IT voltage, time below the serviceability threshold, BESS energy, and converter loading. The apparent-power limit itself is enforced inside the dynamic checker; the resulting peak converter loading is retained as a diagnostic rather than imposed as a separate hard linear constraint in the planning model reported here.

For each data center bus, selected fault, response mode, and target margin $q\in\{V,L\}$, an affine lower-bound surrogate is fitted in the form
\begin{equation}
\hat m^q_{i,t,f}
=
\tilde a^q_{i,f}
+\sum_{j\in\mathcal{C}} b^q_{i,j,f}P^{\mathrm{dc}}_{j,t}
+c^q_{i,f}S^{\mathrm{rated}}_i ,
\label{eq:linear_vrt_surrogate}
\end{equation}
where $\tilde a^q_{i,f}$ is a conservatively tightened intercept. The fitted expression is tightened by the largest training-set overprediction,
\begin{align}
\bar m^{q,(n)}_{i,f}
&=
a^q_{i,f}
+\sum_{j\in\mathcal{C}} b^q_{i,j,f}P^{\mathrm{dc},(n)}_{j}
\notag\\
&\quad
+c^q_{i,f}S^{\mathrm{rated},(n)}_i ,
\label{eq:raw_surrogate_margin}\\
\delta^q_{i,f}
&=
\max_n
\left[
\bar m^{q,(n)}_{i,f}-m^{q,(n)}_{i,f}
\right]_+ .
\label{eq:surrogate_overprediction}
\end{align}

The intercept used in the optimization is then tightened as
\begin{equation}
\tilde a^q_{i,f}=a^q_{i,f}-\delta^q_{i,f},
\label{eq:surrogate_tightening}
\end{equation}

so that the surrogate is conservative on the sampled operating points. With a fixed rating ratio $\beta$, both $P^{\mathrm{dc}}_{j,t}=d_tH_j$ and $S^{\mathrm{rated}}_i=\beta H_i$ are linear functions of the hosting variables. The internal VRT constraints embedded in the hosting problem are therefore
\begin{equation}
\hat m^V_{i,t,f}\geq0,\quad
\hat m^L_{i,t,f}\geq0,
\quad
i\in\mathcal{C},\;t\in\mathcal{T}^{\mathrm{VRT}},\;f\in\mathcal{F}.
\label{eq:embedded_vrt_constraints}
\end{equation}

Equations~\eqref{eq:linear_vrt_surrogate}--\eqref{eq:embedded_vrt_constraints} are the interface between the dynamic VRT checker and the planning optimization. They allow the optimizer to trade capacity among candidate buses while respecting margins learned from network-dependent POI voltage waveforms and internal data center response. Algorithm~\ref{alg:internal_vrt} summarizes this offline-to-optimization workflow. The first part of the algorithm generates training samples by rebuilding the network case, simulating selected faults, and passing the resulting POI voltages through the internal data center model. The second part converts the checked voltage and load-loss margins into tightened affine constraints that can be embedded in the hosting-capacity optimization.

The sampled data are divided into fitting and validation subsets. The tightened model in \eqref{eq:linear_vrt_surrogate} is fitted and tightened only on the fitting subset, while the validation subset is used to report out-of-sample mean absolute error, root-mean-square error, worst overprediction, worst underprediction, and conservative-violation rate for both $m^V$ and $m^L$. To reduce extrapolation risk, the hosting optimization is restricted to the sampled ranges of data center capacity, rating ratio, fault parameters, and representative operating periods used for fitting. After each optimized hosting solution is obtained, the corresponding network case is rebuilt and the complete ANDES fault simulation plus internal data center response model is rerun. A reported hosting solution is accepted only if this nonlinear re-simulation satisfies the sustained-voltage and load-loss criteria for all tested faults and representative VRT periods.

\begin{algorithm}[t]
\caption{Network-coupled internal VRT checking workflow}
\begin{algorithmic}[1]
\State Select candidate buses, operating periods, network faults, and GFL/GFM response modes.
\State Generate sampled hosting vectors around the high-capacity region.
\For{each sample, tested period, and fault}
    \State Build the dynamic simulation case with the sampled hosted data center loads.
    \State Run the disturbance simulation and extract $V^{\mathrm{POI}}_{i,t,f}(\tau)$ at candidate buses.
    \State Drive the internal data center model with the POI waveform.
    \State Compute sustained-voltage and POI load-loss margins.
\EndFor
\State Split samples into fitting and validation sets.
\State Fit and tighten affine surrogate margins for each bus, fault, mode, and target.
\State Evaluate out-of-sample errors and conservative-violation rates.
\State Embed the surrogate constraints in the hosting-capacity optimization.
\State Sweep the rating ratio $\beta$ and GFL/GFM response mode.
\State Re-simulate each optimized solution with the full network and internal dynamic checker.
\end{algorithmic}
\label{alg:internal_vrt}
\end{algorithm}

\section{Case Study}

\begin{figure*}[t]
\centering
\includegraphics[width=0.75\textwidth]{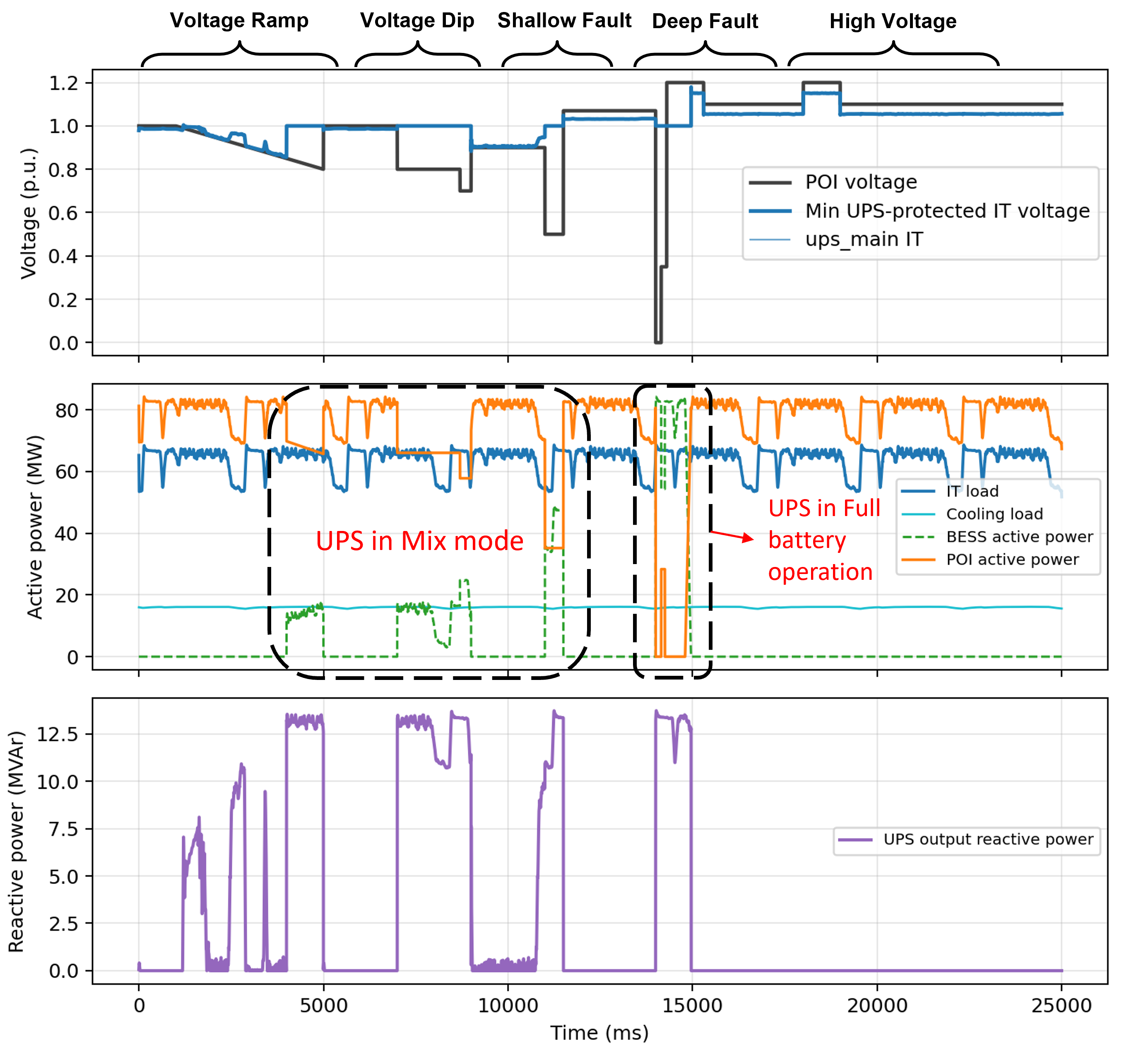}
\caption{Internal data center response under the ERCOT-style POI voltage test profile for the current-limited GFL response mode. The figure shows the imposed POI voltage and protected IT voltage, active-power exchange and BESS supply, and UPS reactive output.}
\label{fig:internal_gfl}
\end{figure*}

\subsection{Data Center Internal Dynamic Response}
We first test the internal ride-through model response under a prescribed POI voltage waveform. The waveform is taken from the updated ERCOT PGRR144 large-load model-quality-test material, specifically the ``Load MQT Testing (POI Voltage)'' curve \cite{ercot2026pgrr144}. It includes voltage-ramp, shallow-fault, deep-fault, and high-voltage segments. This test is used to expose the facility-side dynamics independently from network-fault selection: the POI voltage is imposed as an external disturbance, and the model reports the protected IT-bus voltage, IT and cooling load, POI active power, BESS active supply, and UPS reactive output.

The internal model includes three coupled parts. First, the IT load follows a measured GPU workload trace, scaled with a load-profile gain. Second, the cooling load follows the IT-load multiplier through a first-order thermal response, consistent with the separation between fast IT workload dynamics and slower non-server infrastructure dynamics discussed in recent data center workload studies \cite{ko2026aiworkloads}. Third, the UPS/BESS interface is constrained by the apparent-power headroom of the converter. Fig.~\ref{fig:internal_gfl} illustrates the current-limited GFL response under the prescribed POI waveform and shows the quantities used later by the checker: protected-bus voltage, IT and cooling demand, POI active power, BESS active supply, and UPS reactive output. The GFM mode is not plotted separately here because it uses the same input waveform and diagnostics; its different protected-bus voltage regulation effect is assessed in the network-coupled hosting results.

The VRT test illustrates the physical quantities later used by the hosting-capacity checker. The acceptance criterion is not that the internal voltage must follow or exceed the external VRT envelope. Instead, the checker evaluates whether the protected IT voltage remains above the serviceability threshold for a specified sustained window, while also tracking POI active-power loss and converter loading. These diagnostics are then reused when IEEE 118-bus fault simulations generate the POI voltage waveform for each candidate data center location and hosted capacity.

\subsection{System Configuration}
The numerical study uses the IEEE 118-bus system. The system base is \SI{100}{MVA}; the network has 118 buses and 54 generator buses in the MATPOWER case data \cite{zimmerman2011matpower}. Four PQ buses, 21, 75, 106, and 108, are selected as candidate data center locations. Wind turbines are placed at buses 69, 80, and 89 as hourly profiles. The hosting optimization enforces the 8760-hour power-balance, branch-flow, and normal-operation voltage-screening constraints. The optimization is a linear program for each fixed response mode and rating ratio and is solved with HiGHS \cite{highs2020}. Dynamic network simulations are generated with the ANDES-based Python workflow \cite{cui2020andes}. Reported optimized points are then screened with AC power flow and replayed through the full nonlinear network--data-center dynamic checker before being interpreted as dynamically feasible interconnection plans.

\begin{table*}[t]
\caption{Surrogate-Screened Hosting Capacity With Internal VRT Constraints}
\label{tab:main_results}
\centering
\footnotesize
\begin{tabular}{lrrrrrr}
\toprule
Case & $\beta=S^{\mathrm{rated}}/H$ & Total & Bus 21 & Bus 75 & Bus 106 & Bus 108 \\
 &  & (MW) & (MW) & (MW) & (MW) & (MW) \\
\midrule
No VRT requirement & -- & 391.82 & 31.44 & 153.87 & 102.41 & 104.09 \\
GFL & 1.000 & 107.32 & 0.00 & 50.34 & 19.73 & 37.25 \\
GFL & 1.150 & 227.44 & 0.45 & 126.39 & 32.00 & 68.59 \\
GFL & 1.250 & 344.90 & 16.62 & 155.44 & 104.68 & 68.16 \\
GFL & 1.325 & 391.82 & 31.44 & 153.87 & 102.41 & 104.09 \\
GFL & 1.500 & 391.82 & 31.44 & 153.87 & 102.41 & 104.09 \\
GFM & 1.000 & 109.54 & 0.00 & 69.60 & 0.85 & 39.09 \\
GFM & 1.100 & 254.14 & 21.36 & 45.62 & 103.72 & 83.44 \\
GFM & 1.150 & 385.28 & 24.67 & 154.11 & 102.41 & 104.09 \\
GFM & 1.175 & 391.82 & 31.44 & 153.87 & 102.41 & 104.09 \\
\bottomrule
\end{tabular}
\end{table*}

\begin{figure*}[t]
\centering
\includegraphics[width=0.85\textwidth]{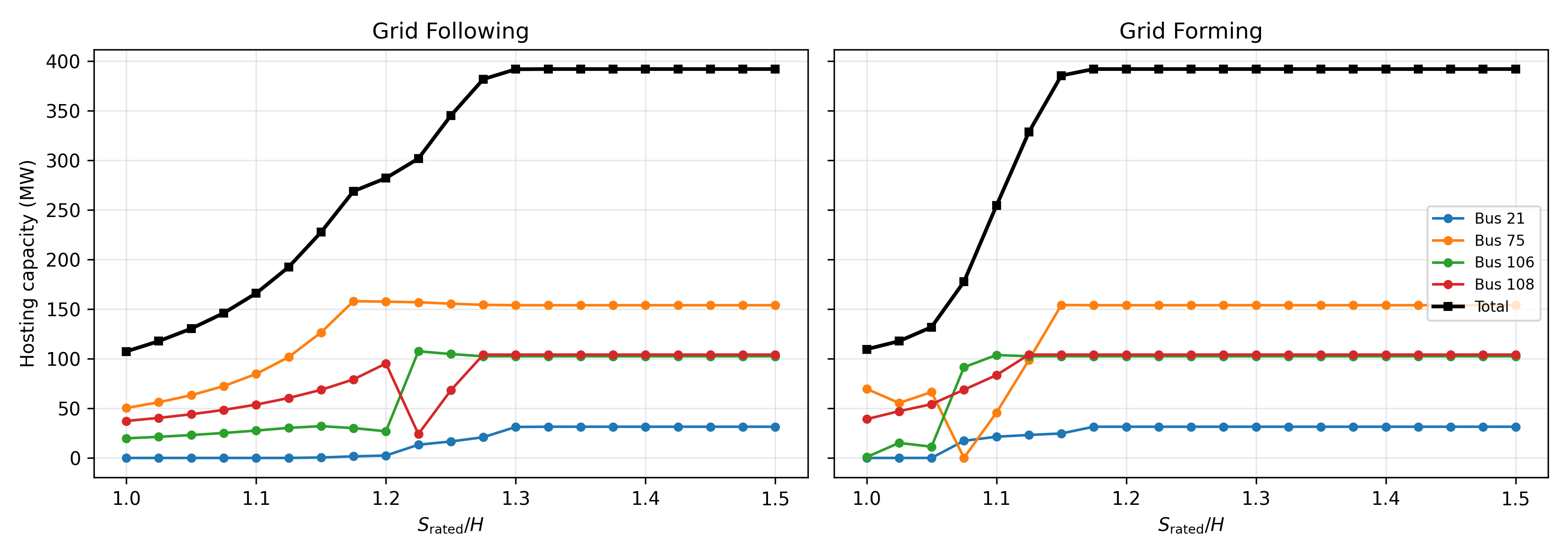}
\caption{Optimized hosting capacity and spatial allocation as a function of UPS/BESS converter rating ratio $\beta=S^{\mathrm{rated}}/H$. Under the reduced-order response models, the GFL case recovers the no-VRT hosting limit at a higher rating ratio than the GFM proxy, and individual bus allocations vary nonmonotonically as the optimizer reallocates capacity under shared network and internal VRT constraints.}
\label{fig:sratio_sweep}
\end{figure*}

The internal VRT training used for the numerical results below is built from selected non-candidate bus faults rather than faults placed directly at the data center buses. Each sampled case rebuilds the IEEE 118 operating point with the sampled data center capacity vector, runs the ANDES fault simulation, and then passes the resulting POI waveform into the internal data center response model. The surrogate data set covers both GFL and GFM response modes, multiple candidate capacity vectors, and rating ratios over the tested VRT capability range. The embedded surrogate is applied to representative VRT operating periods, while the steady-state hosting constraints remain enforced over all 8760 hours. A hosting solution is feasible in the surrogate model if every tested data center bus satisfies the sustained internal-voltage criterion and the maximum POI load loss remains below the specified site-level limit; aggregate POI load loss is recorded for each fault as an additional system-level diagnostic.

\subsection{Hosting Capacity With and Without Internal VRT Constraints}
Table~\ref{tab:main_results} compares the steady-state hosting result with the internal VRT-constrained results obtained from the fitted surrogate. Without internal VRT constraints, the four candidate buses can host \SI{391.82}{MW}. When the current-limited GFL internal response is constrained by a rating ratio of $\beta=1.0$, the feasible hosting capacity falls to \SI{107.32}{MW}. Increasing $\beta$ restores capacity monotonically in the surrogate-screened optimization and reaches the no-VRT allocation near $\beta=1.325$. Under the reduced-order response models used here, the GFM proxy reaches the same surrogate-screened limit at a lower rating ratio, $\beta=1.175$, because it maintains a higher protected-bus voltage margin under the same POI fault waveforms.

\begin{figure}[t]
\centering
\includegraphics[width=0.95\linewidth]{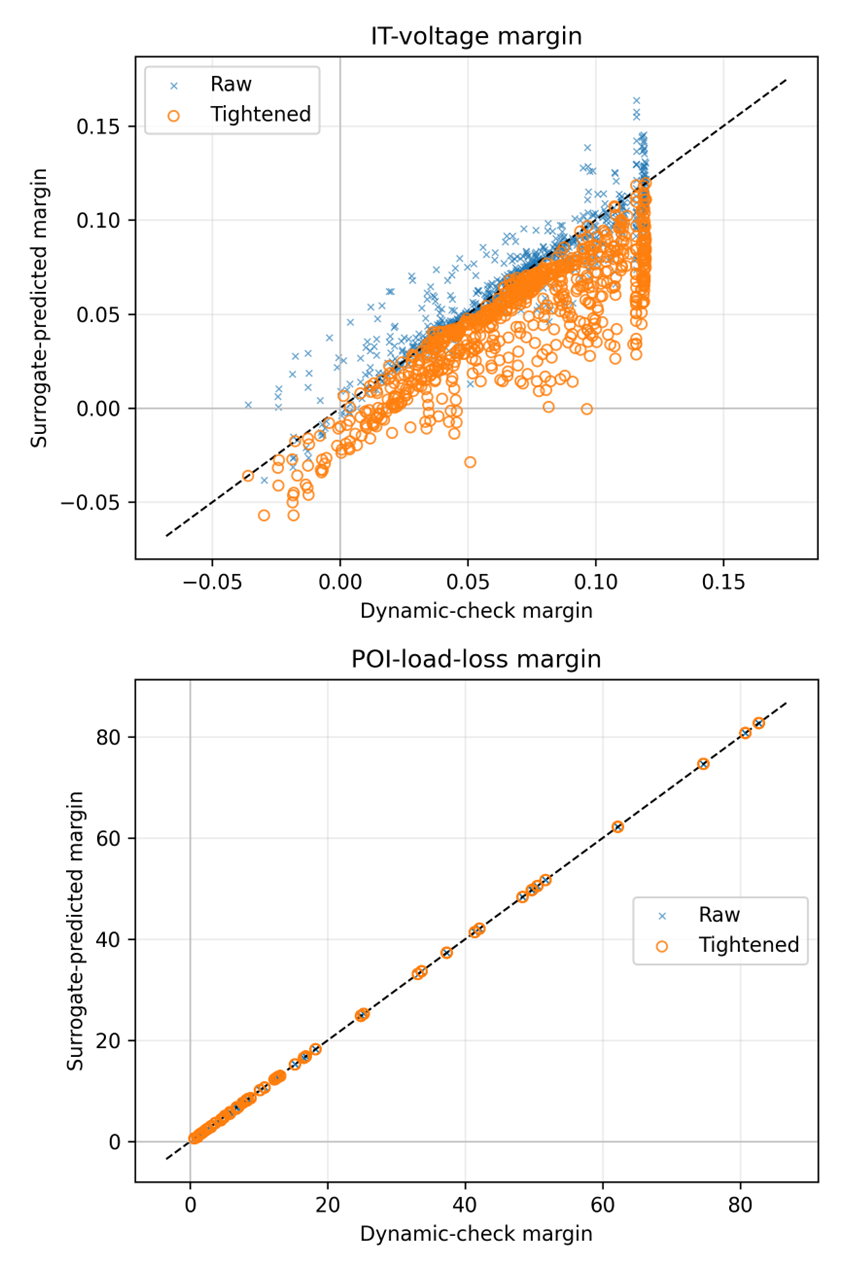}
\caption{Train/test validation of the affine internal-VRT surrogate. The two panels compare dynamic-check margins with surrogate-predicted margins for the internal IT-voltage and POI-load-loss criteria. The tightened surrogate shifts the prediction toward conservative margins.}
\label{fig:surrogate_validation}
\end{figure}

The allocation changes are as important as the total capacity. Bus 21 is initially excluded by the VRT constraints and is gradually unlocked as $\beta$ increases. Bus 75 carries much of the low-capability GFL solution, then falls back toward its no-VRT allocation as buses 106 and 108 become feasible. The non-monotonic allocation at individual buses is expected: the optimizer is not adding independent single-bus VRT capacities, but negotiating among four locations under shared network constraints and fault-dependent internal VRT margins.

Fig.~\ref{fig:sratio_sweep} shows the full rating-ratio sweep for both response modes. The GFL total hosting capacity increases smoothly from \SI{107.32}{MW} at $\beta=1.0$ to the no-VRT limit at $\beta=1.325$. The GFM proxy follows the same qualitative trend but recovers the no-VRT limit earlier, at $\beta=1.175$. This behavior is useful for planning because the result is sensitive to the amount and type of internal VRT capability, not merely to a binary assumption that the data center either can or cannot ride through.

The sweep also illustrates the difference between a local equipment margin and a system-level hosting result. For example, in the GFL case the bus-108 allocation rises to \SI{95.16}{MW} at $\beta=1.20$, drops to \SI{24.05}{MW} at $\beta=1.225$ as capacity shifts toward bus 106, and then returns to the no-VRT allocation as the rating ratio increases further. This behavior is a direct consequence of embedding the fitted VRT margins in the global hosting optimization.

Fig.~\ref{fig:surrogate_validation} summarize the hold-out validation results after residual tightening. The tightened model produces no false-pass cases on the held-out set for either margin. The remaining small positive overpredictions show that the affine screen is conservative in classification but not a formal global lower bound outside the sampled domain.

\section{Conclusion}
This paper proposed a VRT-aware data center--grid co-planning framework that treats internal data center ride-through capability as an explicit interconnection constraint. The method combines steady-state hosting optimization, IEEE 118-bus fault simulation, a reduced-order data center dynamic response model, and a tightened linear surrogate for sustained internal IT voltage and POI load-loss margins. On the IEEE 118-bus case study, the results show that facility-level VRT capability can materially change both the amount and location of feasible data center interconnection. Increasing the UPS/BESS converter rating ratio restores hosting capacity monotonically in the current-limited GFL surrogate sweep, while nonlinear replay shows that additional headroom is needed before the full-capacity GFL solution passes the internal voltage check. Under the same selected network fault conditions, the GFM proxy provides higher ride-through margin and passes the full-capacity replay at a lower rating ratio.

\bibliographystyle{IEEEtran}
\bibliography{references}

\end{document}